\documentclass[12pt,superscriptaddress,nofootinbib,showpacs,showkeys,preprintnumbers]{revtex4}

\usepackage{amsfonts}
\usepackage{mathrsfs}
\usepackage{amssymb}
\usepackage{amsmath}
\usepackage{graphicx}
\usepackage{epstopdf}
\usepackage{pstricks}
\usepackage{slashed}
\usepackage{mathbbol}

\def\be{\begin{equation}}
\def\ee{\end{equation}}
\def\bea{\begin{eqnarray}}
\def\eea{\end{eqnarray}}
\def\bean{\begin{eqnarray*}}
\def\eean{\end{eqnarray*}}
\def\bary{\begin{array}}
\def\eary{\end{array}}
\def\bit{\begin{itemize}}
\def\eit{\end{itemize}}

\def\su5u1{SU(5) \times U(1)}
\def\fsu5u1{SU(5) \times U(1)'}
\def\so10{SO(10)}
\def\sq20{SO(10) \times SO(10)}

\def\bwt{\begin{widetext}}
\def\ewt{\end{widetext}}
\def\be{\begin{equation}}
\def\ee{\end{equation}}
\def\bea{\begin{eqnarray}}
\def\eea{\end{eqnarray}}
\def\bean{\begin{eqnarray*}}
\def\eean{\end{eqnarray*}}
\def\bary{\begin{array}}
\def\eary{\end{array}}
\def\bit{\begin{itemize}}
\def\eit{\end{itemize}}

\def\su5u1{SU(5) \times U(1)}
\def\fsu5u1{SU(5) \times U(1)'}
\def\so10{SO(10)}
\def\sq20{SO(10) \times SO(10)}

\usepackage{amsmath}

\usepackage{amssymb}
\usepackage{gensymb}

\begin{document}
\setlength{\parskip}{0.01cm}


\title{\Large A Symmetric Two Higgs Doublet Model }
\author{Hannah Bossi \footnote{Electronic address: hjbossi@colby.edu},
Shreyashi Chakdar$^a$\footnote{Electronic address: schakdar@colby.edu}}
\affiliation{$^a$Department of Physics and Astronomy, 
Colby College, 5800 Mayflower Hill, Waterville, Maine 04901, USA. \\}
%


\begin{abstract}
\section*{Abstract}
After the discovery of a SM-like scalar state with mass of 125 GeV in 2012, the electroweak sector demands to be studied with all possible theoretical and experimental efforts. In this work we present a symmetric two Higgs doublet model with a discrete interchange symmetry between the two Higgs doublets ($\Phi_1$ $\leftrightarrow$ $\Phi_2$). Apart from the SM-like scalar state (h) with $m_h = 125 $ GeV, the model has several distinguishing features including the pseudoscalar (A), the charged scalars($H^\pm$) and the neutral scalar H, which do not have any direct coupling to fermions. The neutral scalar $H$ can have mass lighter than the 125 GeV SM-like Higgs state $h$. Due to the presence of a residual $Z_2$ symmetry after the Spontaneous Symmetry Breaking (SSB), the neutral scalar $H$ emerges as the Dark Matter candidate in this scenario. As an effect of this possibility, the SM-like scalar $h$ will have an extra invisible decay mode of $h \rightarrow H H$ in this framework. We propose the model and discuss some of the interesting features with a guideline of possible phenomenological searches at the LHC present in this scenario.
\end{abstract}


\maketitle

\section*{Introduction}
With the discovery of a Higgs-like particle $h$ with mass $m_h$ around 125 GeV by the ATLAS and CMS experiments at the LHC \cite{Aad:2012tfa,Chatrchyan:2012ufa}, a new ongoing era of exploration in the electroweak sector began. With this stand-alone SM-like scalar state being observed, it is now vital to explore the scalar sector of the Standard Model in more detail to investigate the possibility of a spectrum of additional scalar states. In this scenario, the simplest and well motivated extension of the Standard Model to be considered are the Two-Higgs-Doublet Models (2HDM's) \cite{Lee:1973iz}, in which adding a second $SU(2)_L$ higgs doublet leads to five physical scalar particles: $h, H, A $ and $H^{\pm}$. If the neutral scalar state $h$ of 2HDM's is considered as the SM-like Higgs state with $m_h \simeq 125$ GeV as seen by the ATLAS and CMS experiments, time seems appropriate to examine all the possible variations of 2HDM's which has the potential to be discovered at the LHC with more data collection at higher energies and luminosities in the near future.\\
There are many motivations behind considering the specific extensions of 2HDMs in Beyond the Standard Model Physics; in Supersymmetric theories \cite{Haber:1984rc}, specifically in the Minimal Supersymmetric Standard Model(MSSM), an additional doublet is needed to give simultaneously masses to the charge 2/3 and charge -1/3 quarks and also for cancellation of the anomalies. Another motivation comes from "axion models" \cite{Kim:1986ax}, in which Peccei and Quinn \cite{Peccei:1977hh} noted that a possible CP-violating term in the QCD Lagrangian can be rotated away if the Lagrangian contains a global U(1) symmetry, which can only be imposed if there are two Higgs doublets. Although the simplest versions of the Peccei-Quinn model are experimentally ruled out now, there are variations that are acceptable which still require two Higgs doublets \cite{Kim:1986ax}. On other hand, one experimental motivation for the 2HDMs is that the SM is unable to generate a sufficient amount of baryon asymmetry (BAO) in the Universe, while due to the flexibility of the mass spectrum and the existence of additional sources of the CP violation, it can be produced in the sufficient amount in the case of 2HDM's \cite{Trodden:1998ym}.\\
Motivated by these features, many versions of the 2HDMs have been extensively studied in the past \cite{Gunion:1989we}. These include: (a) Supersymmetric two Higgs doublet model \cite{Djouadi:2005gj}, (b) Non-supersymmetric two Higgs doublet models:  (i) with only one Higgs doublet (conventionally chosen to be $\Phi_2$) having couplings to the fermions (type I 2HDM), (ii) with both Higgs doublets having VEVs, additionally with one doublet (conventionally chosen to be $\Phi_2$) coupling to the up type quarks only, while the other ($\Phi_1$) coupling to the down type quarks only (type II 2HDM),(iii) with one doublet($\Phi_2$) coupling to the up type quarks only and the other($\Phi_1$) coupling to the down type quarks only i.e just like type II 2HDM along with RH leptons coupling to $\Phi_2$ (``flipped" model or type Y), (iv) with RH quarks coupling to one Higgs doublet($\Phi_2$) and RH leptons coupling to another Higgs doublet($\Phi_1$) (``lepton-specific" or type X) and (v) with only one doublet having VEVs and couplings to the fermions(``Inert Doublet Model"(IDM))\cite{Branco:2011iw}. In another interesting variation of 2HDMs one doublet couples to all the SM fermions except the neutrinos, and has a VEV which is same as the SM VEV($\simeq 250$ GeV), while the other Higgs doublet couples only to the neutrinos with a tiny VEV $\simeq 10^{-2}$ eV \cite{Gabriel:2006ns}. Different variations of all these models have attracted a lot of attention recently and were investigated in view of the recent LHC data \cite{Maitra:2014qea,Altmannshofer:2012ar,Chang:2012ve,Chen:2013kt,Celis:2013rcs,Grinstein:2013npa,Coleppa:2013dya,Chen:2013rba,Eberhardt:2013uba,  Craig:2013hca,Maiani:2013nga,Wang:2013sha,Baglio:2014nea,Kanemura:2014dea,Dumont:2014wha}

In this work, we present a symmetric two Higgs doublet model with a discrete interchange symmetry ($\Phi_1$ $\leftrightarrow$ $\Phi_2$) having several distinguishing features that include:\\
(i) The neutral scalar $h$ behaves like the recently discovered Standard Model(SM)-like Higgs with mass $m_h \simeq 125$ GeV.\\
(ii) The model has no flavor-changing neutral currents (FCNC).\\
(iii) The other neutral scalar $H$ can be lighter than the SM-like neutral scalar state $h$ and couples only to the Electroweak Gauge Bosons ($W^\pm$ and $Z$) but does not couple to fermions at all.\\
(iv) The charged Higgs $H^{\pm}$ and the pseudoscalar $A$ also do not couple to the fermions. The charged scalar only couples to the gauge bosons, the neutral scalar $H$ and pseudoscalar $A$.\\
(v) The pseudoscalar only couples to the gauge bosons and the charged scalar $H^{\pm}$ but does not couple to either the lighter and heavier neutral scalars ($H$ and $h$).\\
(vi) At the LHC, the lighter scalar $H$ can only be produced via the decays of the SM-like neutral scalar $h$. As the lighter scalar $H$ does not couple to the fermions at all, the mass limit on this neutral scalar will be significantly lower than the mass limits coming from LHC data using gluon-gluon fusion process for production of a generic scalar boson.\\
(vii) The charged Higgs $H^{\pm}$ can be produced via the Drell-Yan process. It will be quite elusive to discover it at the LHC with the decay products of  "2 W-boson + Miss $E_T$" in the final state.\\
(viii) After the $\Phi_1$ $\leftrightarrow$ $\Phi_2$ interchange symmetry is spontaneously broken, there is a residual $Z_2$ symmetry that remains unbroken. This residual symmetry makes the scalar states $H^{\pm} $, $H$ and $ A$  to become $Z_2$ negative, while all other fields become $Z_2$ positive. 
Thus the lightest $Z_2$ negative particle (either $H$ or $A$ ) will be a candidate for the dark matter (DM) in this scenario.\\
(ix) If the lightest scalar $H$ is taken to be a DM candidate, the decay channel $h \to H H$ will act as an extra invisible decay mode of the SM-like Higgs with $m_h \simeq 125$ GeV. We can get a bound in the parameter space of the effective coupling between the two neutral scalars $h$ and $H$ and the mass of the lightest scalar $m_H$ coming from the invisible Higgs branching ratio bound for the SM-like state: $BR_{inv h} < 25\%$.\\
(x) We also investigate the various bounds on the dark matter parameter space; for example, bounds from Electroweak precision constraints and Direct and Indirect detection are discussed.\\
(xi) The neutral Higgs $H$ has a coupling $H H Z Z$  which can lead to interesting decay process $Z \to Z^* H H \to f \bar{f} H H$, that can be tested at the proposed $e^+e^-$ collider ILC. \cite{Djouadi:2007ik}\\
 Below we present the Model and formalism, followed by the phenomenological implications for this ``Symmetric two Higgs doublet model."

\section*{The Model and formalism: ``A Symmetric two Higgs doublet model"}
Our proposed model is based on the Standard Model (SM) symmetry group $SU(3)_C \times SU(2)_L \times U(1)_Y$ with the Higgs sector being extended minimally with two Higgs doublets $\Phi_1$ and $\Phi_2$. We consider a discrete interchange symmetry of $\Phi_1$ $\leftrightarrow$ $\Phi_2$ with the condition of $V_{\Phi_1} = V_{\Phi_2}$. Thus the proposed model is named "A symmetric two Higgs doublet model".\\
The VEV's of these Higgs doublets are summarized below:
\begin{eqnarray}
\left < \Phi_1 \right> = \frac{1}{\sqrt 2}{\begin{pmatrix} 0 \\ V_{\Phi_1} \end{pmatrix}} &,& \left < {\Phi_2} \right> = \frac{1}{\sqrt 2}{\begin{pmatrix} 0 \\ V_{\Phi_2} \end{pmatrix}}
 \label{vev}.
\end{eqnarray} 
In the unitary gauge, the two doublets can be written as,  
\begin{eqnarray}
\Phi_1  =\frac{1}{\sqrt 2} {\begin{pmatrix}  \frac{\sqrt 2 V_{\Phi_2}}{V} H^+ \\ h_0 + i\frac{V_{\Phi_2}}{V} A + V_{\Phi_1}  \end{pmatrix}} &,& \Phi_2  = \frac{1}{\sqrt 2} {\begin{pmatrix}  \frac{-\sqrt 2 V_{\Phi_1}}{V} H^+ \\ H_0 - i\frac{V_{\Phi_1}}{V} A + V_{\Phi_2}  \end{pmatrix}}
 \end{eqnarray}
 where $V_{\Phi_1} = V_{\Phi_2} = v/\sqrt 2$  and $v^2 = V_{\Phi_1}^2 + V_{\Phi_1}^2 = (250)^2$ GeV. The five physical scalar fields are $h , H, A$  and $H^{\pm}$ which are respectively the two neutral scalars, the psuedoscalar and the charged scalar. After the $\Phi_1$ $\leftrightarrow$ $\Phi_2$ interchange symmetry is spontaneously broken, there is a residual $Z_2$ symmetry that remains unbroken. This residual symmetry makes $H^{\pm},H$ and $A$ acquire $Z_2$ negative charges i.e $H^{\pm} \rightarrow - H^{\pm}$ ;  $H \rightarrow -  H$ ;  $A \rightarrow -A$ while all other fields are $Z_2$ positive. As a result, the lightest $Z_2$ negative particle ( either $H$ or A ) could be considered as a suitable candidate for the dark matter(DM). 

 The most general potential with this exchange symmetry of $\Phi_1$ $\leftrightarrow$ $\Phi_2$ can be written as,\\
\begin{eqnarray}
V=&+&m_{11}^2 \Phi_1^\dagger \Phi_1 + m_{22}^2\Phi_2^\dagger \Phi_2  - m_{12}^2 \Phi_1^\dagger \Phi_2 - m_{12}^{*2}\Phi_2^\dagger \Phi_1 + \frac{\lambda_1}{2} \left(\Phi_1^\dagger \Phi_1 \right)^2 +\frac{\lambda_2}{2}\left(\Phi_2^\dagger \Phi_2 \right)^2 + \lambda_3 \left(\Phi_1^\dagger \Phi_1 \right) \left(\Phi_2^\dagger \Phi_2 \right)\nonumber\\
&+&\lambda_4 \left(\Phi_1^\dagger \Phi_2 \right) \left(\Phi_2^\dagger \Phi_1 \right)+\frac{\lambda_5}{2} \left(\Phi_1^\dagger \Phi_2 \right)^2 + \frac{\lambda^*_5}{2} \left(\Phi_2^\dagger \Phi_1 \right)^2 + \lambda_6 \left(\Phi_1^\dagger \Phi_1 \right) \left(\Phi_1^\dagger \Phi_2 \right)+ \lambda_6^* \left(\Phi_1^\dagger \Phi_1 \right) \left(\Phi_2^\dagger \Phi_1 \right)\nonumber\\
&+&\lambda_7 \left(\Phi_2^\dagger \Phi_2 \right) \left(\Phi_1^\dagger \Phi_2 \right) + \lambda_7^* \left(\Phi_2^\dagger \Phi_2 \right) \left(\Phi_2^\dagger \Phi_1 \right)
\label{potential}
\end{eqnarray}
where due to the exchange symmetry of $\Phi_1$ $\leftrightarrow$ $\Phi_2$, we take,
\begin{eqnarray}
m_{11}^2 = m_{22}^2\nonumber\\
m_{12}^2 = m_{12}^{*2}\nonumber\\
\lambda_1 = \lambda_2\nonumber\\
\lambda_5 = \lambda_5^*\nonumber\\
\lambda_6 = \lambda_7^*\nonumber
\end{eqnarray}
so that we have total 8 parameters in the potential.
Minimization of this potential gives us two solutions,
\begin{eqnarray}
& & V_{\Phi_1} = V_{\Phi_2} ,
\label{min}
\end{eqnarray}
\begin{eqnarray}
&&m_{11}^2 + m_{12}^2 + \lambda_1(V_{\Phi_1}^2 +V_{\Phi_2}^2 + V_{\Phi_1}V_{\Phi_2}) - (\lambda_3 + \lambda_4 + \lambda_5)V_{\Phi_1}V_{\Phi_2} - \lambda_6(V_{\Phi_1}^2 + V_{\Phi_2}^2) +2\lambda_6^* V_{\Phi_1}V_{\Phi_2} = 0 \nonumber\\
\end{eqnarray}
among which we choose the first solution of minimization i.e $V_{\Phi_1} = V_{\Phi_2} = v/\sqrt 2 = 175$ GeV. 
From the determination of the second derivatives of the potential V referred in Eqn(\ref{potential}), the first minimization solution of Eqn(\ref{min}) can be taken as a local minima along with the following condition,
\begin{eqnarray}
(\lambda_1  + \lambda_5 +2[\lambda_6^*]) \geq 0
\label{higgscond}
\end{eqnarray}
In this work, we consider the first minimisation solution with $V_{\Phi_1} = V_{\Phi_2} = v/\sqrt 2 = 175$ GeV along with the condition stated in Eqn(\ref{higgscond}) above. 
With the above minimum, the particle masses are given by,
\begin{eqnarray}
m_w^2 = \frac{g^2v^2}{4}  , m_Z^2 = \frac{(g^2 + g'^2)v^2}{4},\nonumber \\
 m_{h}^2 = (\lambda_3 + \lambda_4 + \lambda_5 - \lambda_1)v^2/2 , \nonumber\\
m_{H}^2 = 2m_{12}^2 - [\lambda_1 + \lambda_3 + \lambda_4 + \lambda_5 + 2(\lambda_6 + \lambda_6^*)]v^2/2 ,\nonumber\\
m_{A}^2 = 2m_{12}^2 - [\lambda_5 + \lambda_6 + \lambda_6^* ]v^2 ,\nonumber\\
m_{H^{\pm}}^2 = 2m_{12}^2 - [ \lambda_4 + \lambda_5 + \lambda_6 + \lambda_6^*]v^2 , 
\end{eqnarray}
\\
where the two neutral scalars $h$ and $H$ are defined as,
\begin{eqnarray}
h =  h^0 cos(\alpha) + H^0 sin(\alpha),\nonumber\\
H =  - h^0 sin(\alpha) + H^0 cos(\alpha)
\label{orthogonalisation}
\end{eqnarray}
where $\alpha$ = the scalar diagonalising angle = $\pi/4$. Also,
here $tan\beta = V_{\Phi_2}/V_{\Phi_1}= 1$.

From the assumption that we take the neutral scalar $H$ to be lighter that the SM-like neutral scalar $h$ and $m_{H}^2 < m_{h}^2$, we get a condition on the linear combination of the parameters as below
\begin{eqnarray}
m_{12}^2 < [\lambda_3 + \lambda_4 + \lambda_5 + \lambda_6 + \lambda_6^*]v^2/2
\end{eqnarray}

Now as the neutral scalar $h$ in our model is considered as the Higgs state (SM-like) with $m_h \simeq 125$ GeV as seen by the LHC, we get another condition on the relevant parameters given by
\begin{eqnarray}
\lambda_3 + \lambda_4 + \lambda_5 - \lambda_1 = 0.25 
\label{mh cond}
\end{eqnarray} 

Also, the potential must be bounded from below and the corresponding vacuum stability conditions are given by,
\begin{eqnarray}
\sqrt{4\pi} > \lambda_1 \geq 0 ,
\sqrt{4\pi} > \lambda_2 \geq 0 ,\nonumber\\
\lambda_3 \geq - \sqrt{\lambda_1\lambda_2} ,
\lambda_3  + \lambda_4 - [\lambda_5] \geq - \sqrt{\lambda_1\lambda_2},
\label{boundl}
\end{eqnarray}
Note that these conditions are only valid when $\lambda_6 = \lambda_7 = 0$.\\

Furthermore, from the LEP experiment we have constraints on the Z boson decay width and mass of the charged scalar ($H^\pm$) \cite{Beringer:1900zz} given by,
\begin{eqnarray} 
m_H + m_A > m_Z,\nonumber \\
m_H^\pm > 79.3  \textnormal {GeV}.
\end{eqnarray}


If we now look at the Yukawa interaction Lagrangian relevant under our gauge symmetry as well as the discrete interchange symmetry of $\Phi_1$ $\leftrightarrow$ $\Phi_2$ given by,
\begin{eqnarray}
{\cal L}  &\supset& f_d^{\Phi_1} (\bar{u_L},\bar{d_L})d_R \Phi_1 + f_u^{\Phi_1} (\bar{u_L},\bar{d_L})u_R \tilde{\Phi_1} 
+ f_d^{\Phi_2} (\bar{u_L},\bar{d_L})d_R \Phi_2 + f_u^{\Phi_2} (\bar{u_L},\bar{d_L})u_R \tilde{\Phi_2} ,\nonumber\\
\end{eqnarray}  
where because of the interchange symmetry $\Phi_1$ $\leftrightarrow$ $\Phi_2$, we take
\begin{eqnarray}
f_d^{\Phi_1} = f_d^{\Phi_2} = f_d \nonumber\\
f_u^{\Phi_1} = f_u^{\Phi_2} = f_u 
\end{eqnarray}
from this we can obtain the charged scalar ($H^{\pm}$) couplings, the neutral scalar ($h, H$) couplings and the pseudoscalar ($A$) couplings with the fermions. We find that only the neutral scalar ($h$) couple to the fermions with the couplings of  $f_u$ and $f_d$(same as in the Standard Model (SM)) and the other scalars ($H, H^{\pm},A$) do not couple to the fermions at all.\\

Again, looking at the gauge bosons masses and mixings are obtained from the kinetic terms of the scalars in the Lagrangian, we get
\begin{equation}
{\cal L} \supset \left({\cal D}_\mu\Phi_1\right)^\dagger\left({\cal D}^\mu\Phi_1\right)+\left({\cal D}_\mu\Phi_2\right)^\dagger\left({\cal D}^\mu\Phi_2\right),\label{ktl}
\end{equation}
where, ${\cal D}$ is the covariant derivative associated with the gauge group, given by
 {\begin{equation}
{\cal D}_\mu =\partial_\mu - ig \frac{\tau_a}{2}A^a_\mu - ig^\prime \frac{Y}{2} B_\mu,
\end{equation}}
where $\tau_a$'s are the Pauli matrices. The relevant interactions of the scalar fields $h^0, H^0, H^{\pm}$ and $A$ with the gauge bosons ( $W^\pm$ and Z) before the spontaneous symmetry breaking (SSB) are given by,
\begin{eqnarray}
{\cal L}_{gauge}  &\supset& gm_w W_{\mu}^+W_{\mu}^-\frac{h^0 + H^0}{\sqrt 2} + \frac{m_Z^2}{v} Z_{\mu}Z_{\mu}\frac{h^0 + H^0}{\sqrt 2} + \frac{m_w^2}{v^2} W_{\mu}^+W_{\mu}^-((h^0)^2 + (H^0)^2) \nonumber\\
&+& \frac{m_Z^2}{2v^2} Z_{\mu}Z_{\mu}((h^0)^2 + (H^0)^2) - \frac{g}{2} (\partial_\mu H^\pm)W_{\mu}^\pm\frac{H^0 - h^0}{\sqrt 2} - \frac{g}{2} \partial_\mu (\frac{H^0 - h^0}{\sqrt 2}) W_{\mu}^\pm H^\pm \nonumber\\
&\pm& \frac{ig}{2}(\partial_\mu H^\mp)W_{\mu}^\pm A \pm \frac{ig}{2}(\partial_\mu A)W_{\mu}^\pm H^\mp
\label{Lgauge}
\end{eqnarray} 

Now with the definition of the mass eigenstates $h$ and $H$ as stated in eqn(\ref{orthogonalisation}) with  $\alpha$ = the scalar diagonalizing angle = $\pi/4$ we get,
\begin{eqnarray}
h =  (h^0  + H^0) / \sqrt 2,\nonumber\\
H =  (- h^0  + H^0)/ \sqrt 2
\label{scalar def}
\end{eqnarray}
Using these definitions of Eqn(\ref{scalar def}) in Eqn(\ref{Lgauge}) we get the relevant interactions after the SSB as,
\begin{eqnarray}
{\cal L}_{gauge}  &\supset& gm_w W_{\mu}^+W_{\mu}^- h + \frac{m_Z^2}{v} Z_{\mu}Z_{\mu} h + \frac{m_w^2}{v^2} W_{\mu}^+W_{\mu}^- h^2 + \frac{m_w^2}{v^2} W_{\mu}^+W_{\mu}^- H^2 \nonumber\\
&+& \frac{m_Z^2}{2v^2} Z_{\mu}Z_{\mu}h^2 + \frac{m_Z^2}{2v^2} Z_{\mu}Z_{\mu}H^2 - \frac{g}{2} (\partial_\mu H^\pm)W_{\mu}^\pm H - \frac{g}{2} (\partial_\mu H) W_{\mu}^\pm H^\pm \nonumber\\
&\pm& \frac{ig}{2}(\partial_\mu H^\mp)W_{\mu}^\pm A \pm \frac{ig}{2}(\partial_\mu A)W_{\mu}^\pm H^\mp
\label{L_gauge}
\end{eqnarray} 
Interestingly, as seen from the above interactions, only the SM-like scalar $h$ has three point couplings with the gauge bosons ($W^\pm$ and Z) that are same with the couplings found in the Standard Model(SM), where as the other neutral scalar $H$ does not have any three point couplings of this sort. However, the neutral scalar $H$ has four point couplings with the gauge bosons ($W_{\mu}^+W_{\mu}^- H H $ and $Z_{\mu} Z_{\mu} H H$). The charged scalars ($H^{\pm}$) couple to both the neutral Higgs $H$ and the pseudoscalar $A$ along with the gauge boson $W^\pm$ whereas the pseudoscalar $A$ does not couple to the neutral scalar state $H$ but only couples to the charged scalar ($H^{\pm}$) and the gauge boson $W^\pm$.\\
From the Higgs potential, we get these relevant interactions of the scalar fields $h^0, H^0, A$ and $H^{\pm}$ before the SSB,
\begin{eqnarray}
{\cal L}_{scalar}  &\supset& \frac{(\lambda_3 + \lambda_4 + \lambda_ 5)}{4} ((h^0)^2(H^0)^2 + \sqrt 2 v ((h^0)^2H^0 +(H^0)^2h^0 ))\nonumber\\ &+& \frac{(\lambda_6 + \lambda_6^*)}{4}(((h^0)^3H^0 +(H^0)^3h^0) + \frac{3v((h^0)^2H^0 +(H^0)^2h^0 )}{\sqrt2})
\end{eqnarray} 
After the SSB the relevant interactions between the two neutral scalars $h$ and $H$ are given by,
\begin{eqnarray}
{\cal L}_{scalar}  &\supset& -\frac{(\lambda_3 + \lambda_4 + \lambda_ 5)}{4} h^2 H^2 - (\frac{2(\lambda_3 + \lambda_4 + \lambda_ 5)+3(\lambda_6 + \lambda_6^*)}{8}) v h H^2  
\label{scalar coup}
\end{eqnarray}
So it can be seen from the interactions that there is a four point coupling $hhHH$ and a three point coupling $hHH$ between the two neutral scalars $h$ and $H$.\\

As discussed before, after the $\Phi_1$ $\leftrightarrow$ $\Phi_2$ interchange symmetry is spontaneously broken, there is a residual $Z_2$ symmetry that can be considered to remain unbroken. This residual symmetry makes $H^{\pm}, H$ and $A$ to acquire $Z_2$ negative charges i.e $H^{\pm} \rightarrow - H^{\pm}$ ;  $H \rightarrow -  H$ ;  $A \rightarrow -A$, while all other fields acquire $Z_2$ positive charges. 
Thus the lightest $Z_2$ negative particle (either the neutral scalar $H$ or the pseudoscalar $A$ ) can be considered to be a dark matter(DM) candidate in our model.\\
 In this model the masses of the neutral scalar $H$ and the pseudoscalar $A$ are given by,\\
$m_{H}^2 = 2m_{12}^2 - [\lambda_1 + \lambda_3 + \lambda_4 + \lambda_5 + 2(\lambda_6 + \lambda_6^*)]v^2/2 $,\nonumber\\
$m_{A}^2 = 2m_{12}^2 - [\lambda_5 + \lambda_6 + \lambda_6^* ]v^2 $,\nonumber\\
Assuming $m_{H}^2 < m_{A}^2$ and using the condition referred in Eqn(\ref{mh cond}) and using the bound on $\lambda_1$ stated in Eqn(\ref{boundl}), we get a bound on $\lambda_5$ for the lighter neutral scalar $H$ to become the lightest $Z_2$ negative particle and fulfill the role of the DM candidate in this model:
\begin{eqnarray}
3.67 > \lambda_5 \geq 0.125
\end{eqnarray}
Hence the neutral scalar $H$ can serve as the Dark Matter candidate (DM) as it will be the lightest scalar particle protected by the residual $Z_2$ symmetry. This DM candidate $H$ must also satisfy the correct relic abundance of DM obtained from the PLANCK Collaboration \cite{Ade:2013zuv} as well as the electroweak precision constraint bounds \cite{Hook:2011}, dark matter direct detection \cite{Aprile:2013} and indirect detection \cite{Elor:2016} bounds given by,
\begin{eqnarray}
\Omega_{DM}h^2 = 0.1199 \pm 0.0027,
\end{eqnarray}
where $\Omega$ is the density parameter and $h$ is the Hubble parameter in the unit of 100 km $s^{-1}$ Mp$c^{-1}$. 

\section*{Phenomenological implication}
In this section, we consider the phenomenological implications of this model. There are several interesting phenomenological implications which can be tested in the next runs of LHC at 14 TeV and the proposed $e^+e^-$ collider ILC \cite{Djouadi:2007ik}. As the neutral scalar boson $h$ has the same couplings as the Higgs boson in the Standard Model, the production and decays of the neutral scalar ($h$) particle will be same as of the SM-like Higgs state with $m_h \simeq 125$ GeV as seen by the LHC. Interestingly, all the other scalars ($H, H^{\pm}, A$) do not couple to the fermions. For the other neutral scalar $H$, assuming that its mass is lighter that the SM-like Higgs boson mass ($m_H < m_h$), this neutral scalar can be produced via the decays of the SM-like Higgs $h$ having mass of $125$ GeV, through the three point coupling between the two neutral scalars $h$ and $H$ of the form $hHH$, as seen in equation(\ref{scalar coup}). As stated previously, as the neutral scalar $H$ is the lightest stable $Z_2$ negative particle in the scalar particle mass spectrum, it can not decay to anything further. So the decay of the $125$GeV Higgs $h$ into two lighter scalars $H$ will be seen as invisible decays in the detectors. As a result, there will be an extra invisible decay mode for the SM-like scalar $h$ via the decay mode $h \rightarrow H H$.  The Invisible decay branching ratio of the $125$ GeV SM-like Higgs can be as large as $Br_{inv} < 25 \%$ \cite{Banerjee:2017kya}.
The expression for the partial decay width of the SM-like Higgs boson $h$ with mass of $m_h \simeq 125$ GeV to two lighter neutral scalar particles $H$ is given by,
\begin{eqnarray}
\Gamma(h \to H H)&=&\frac{v^2}{2048 \pi M_h}\left(2(\lambda_3 +\lambda_4+\lambda_5) + 3(\lambda_6+\lambda_6^*) \right)^{2}(1- \frac{4m_H^2}{m_h^2})^{1/2}
\label{hHH}
\end{eqnarray}  
\\
Note that the total width of the SM-like Higgs boson with mass of $125$ GeV can be expressed in our model as,
\begin{eqnarray}
\Gamma_{h}^{tot}&=& \sum\Gamma_{h \rightarrow A\bar{A}} + \Gamma_{h \to H H}
\end{eqnarray}
\begin{figure}
\begin{center}
\includegraphics[scale = 0.7]{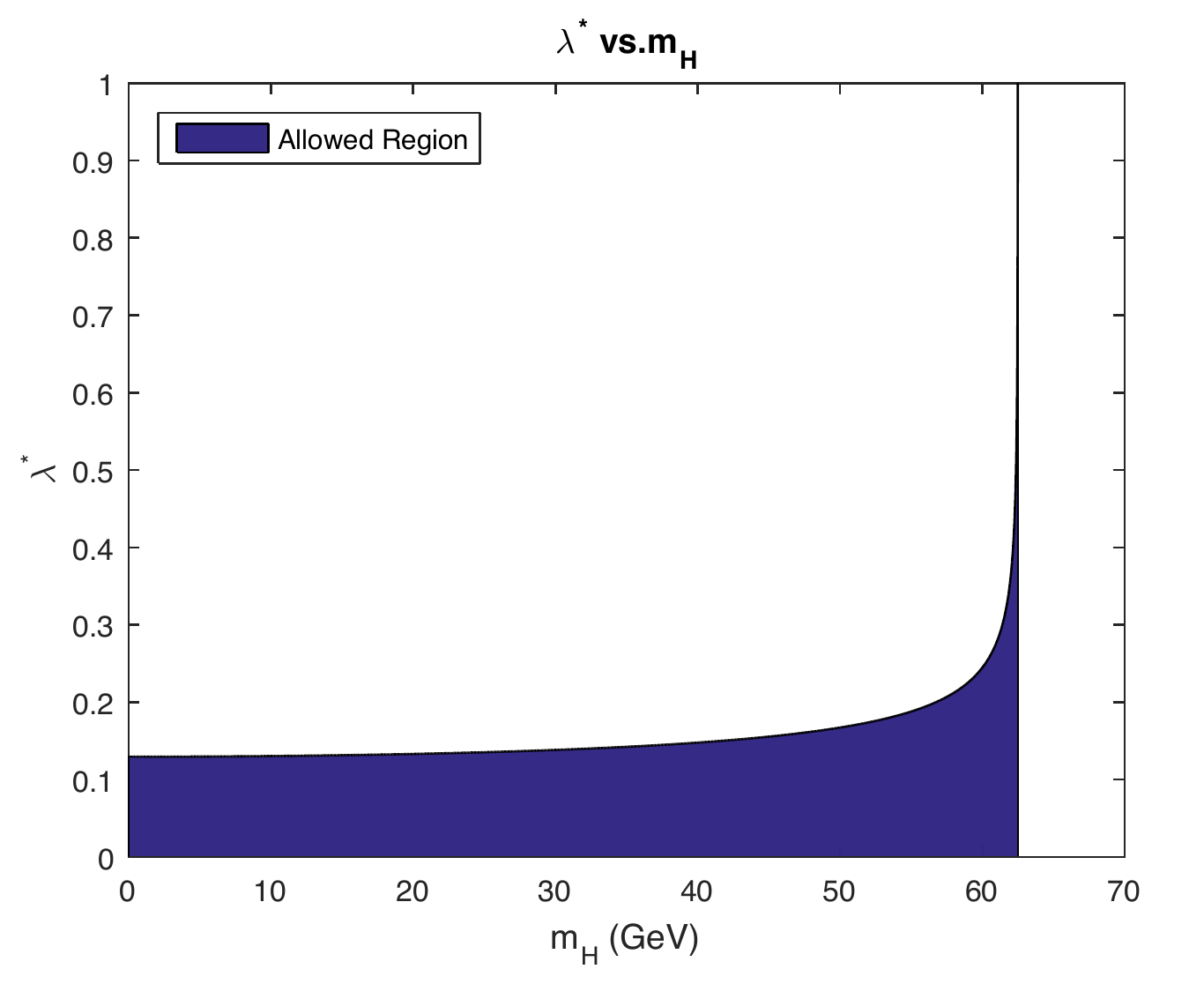}
\caption{Variation of the effective coupling $\lambda^*$ vs mass of the DM ($m_H$) in the symmetric two Higgs doublet model is shown in this figure. The shaded region corresponds to the allowed region in the parameter space for $\lambda^*$ and $m_H$ coming from the current bound of $125$ GeV SM-like Higgs invisible decay Branching Ratio $ < 25\%$.} 
\label{fig:H_lambda}
\end{center}
\end{figure}
where $A\bar{A} = b\bar{b}, \tau\tau, gg, WW^*, ZZ^*, c\bar{c}, \gamma\gamma$ and $\sum\Gamma_{h \rightarrow A\bar{A}}$ is taken to be the total decay width of a $125$ GeV SM-like Higgs boson $= 4.088$ MeV \cite{Denner:2011mq}. The partial decay width of $\Gamma(h \to H H)$ adds to the invisible decay branching ratio of the $125$ GeV SM-like Higgs. So from the upper bound of ${Br_{inv}}_h < 25 \%$ \cite{Banerjee:2017kya} for the invisible deacy Branching Ratio of the SM-like Higgs particle, we get the following bound on the decay width ($\Gamma(h \to H H)$)\\
\begin{eqnarray}
\Gamma(h \to H H) < 1.2775 \times 10^{-3}  \textnormal {GeV}
\end{eqnarray}
Using Eqn(\ref{hHH}) this can be translated as a bound on the effective coupling $\lambda^*$ and $m_H$ as,
\begin{eqnarray}
{\lambda^*}^2(1- \frac{4m_H^2}{m_h^2})^{1/2} < 0.01678
\end{eqnarray}
where $\lambda^*$ = $2(\lambda_3 +\lambda_4+\lambda_5) + 3(\lambda_6+\lambda_6^*)$ is defined to be the effective coupling between $hHH$.

The variation of $\lambda^*$ vs $m_H$ is shown in Fig(\ref{fig:H_lambda}). The shaded region corresponds to the allowed parameter space from the invisible decay branching ratio bound of $ {Br_{inv}}_h < 25\%$ of the $125$ GeV Higgs boson. It can be seen in Fig(\ref{fig:H_lambda}) that the effective coupling $\lambda^*$ seems to be varying very slowly with $m_H$ for the lower neutral scalar mass($m_H$) values while at $m_H = 62.5$, the effective coupling $\lambda^*$ becomes infinity. 

\begin{figure}[ht]
	\centering
	\includegraphics[scale = 0.4]{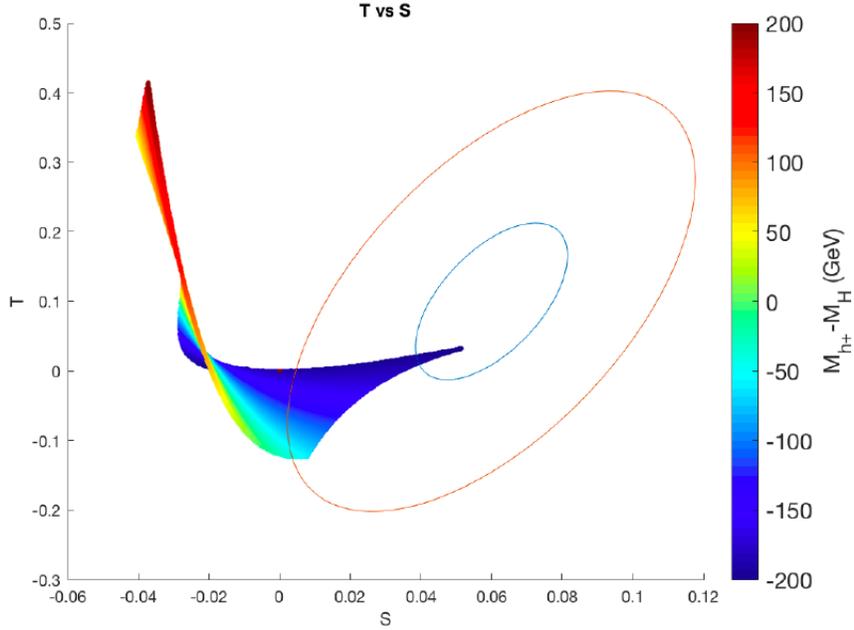}
	\caption{Effect of the S-T Constraint on the $M_{h+} - M_{h2}$ mass split. Figure shows the color map of the $M_{h+} - M_{h2}$ mass split in the S-T plane together with the 1-$\sigma$ (blue ellipse) and $2\sigma$ $\chi^{2}$ (red ellipse) confidence intervals calculated with two degrees of freedom.}
	\label{STfig}
\end{figure}

So, from the plot we get a glimpse of the parameter space for allowed values of $\lambda^*$ and $m_H$ in this model. Note that the bounds which are reflected in Figure \ref{fig:H_lambda} only reflect to the invisible Higgs decay LHC bounds. In order to study the parameter space in detail, one needs to consider other bounds such as direct and indirect detection bounds along with electroweak precision constraint bounds. Some of these bounds are considered in Figure \ref{STfig}.\\
While studying the phenomenology of the symmetric 2HDM, we wanted to test the bounds coming from the  Electroweak Precision Tests (EWPT). As we know, EWPT can be expressed in terms of
three measurable quantities, called S, T, and U, that parameterize contributions from beyond
standard model physics to electroweak radiative corrections \cite{Peskin:1991sw}. The contribution to the S and T parameters \cite{Hook:2011} in this framework can be written as
\begin{eqnarray}
 S &=&\frac{1}{72 \pi (x_2^2 -x_1^2)^3} [x_2^6 f_a(x_2) - x_1^6 f_a(x_1) + 9 x_2^2 x_1^2 (x_2^2 f_b(x_2) - x_1^2 f_b(x_1))] 
 \end{eqnarray}
 where, 
 \begin{eqnarray}
 x_1 = \frac{m_{h}}{m_{h+}}, x_2 = \frac{m_{H}}{m_{h+}}, 
 f_a = -5 + 12log(x),  f_b = 3 - 4log(x)
\end{eqnarray}
and
\begin{eqnarray}
 T &=&\frac{1}{32 \pi^2 \alpha v^2} [f_c(m_{h+}^2, m_{H}^2) + f_c(m_{h+}^2, m_{h}^2) - f_c(m_{H}^2, m_{h}^2)] 
 \end{eqnarray}
 where,
\begin{eqnarray}
f_c = \frac {x+y}{2} - \frac {xy}{x-y}log(\frac{x}{y}), x \neq y\\
f_c = 0, x =y
\end{eqnarray}
With U fixed to be zero, the central values of S and T, assuming a SM Higgs boson mass of $m_h$ = 125 GeV, are given by \cite{Baak:2014ora}:
\begin{eqnarray}
S = 0.06 \pm 0.09, T = 0.1 \pm 0.07
\end{eqnarray}

In Figure \ref{STfig}, we can see the allowed parameter space in this framework with respect to the Electroweak precision bounds (S, T). The Figure \ref{STfig} represents the color map of the
($m_{h+} - m_{H}$) mass split in the (S; T) plane together with the 1-$\sigma$ and $2\sigma$ $\chi^{2}$  confidence intervals calculated with two degrees of freedom. One can see that EWPT data prefer a modest negative ($m_{h+} - m_{H}$) mass split below about 100 GeV, which is due to the roles and respective range of variation for the following S and T parameter bound. 

The DM candidate lighter neutral scalar $H$ can also be produced through the unsuppressed 4-point coupling $Z Z H H$ as seen in Eqn(\ref{L_gauge}). We consider the decay process $Z \rightarrow Z^* H H \rightarrow f\bar{f} H H$. 
The decay width with an assumption of $m_H = 0$ gives \cite{Gabriel:2006ns}
\begin{eqnarray}
\sum_f \Gamma(Z \rightarrow f\bar{f} H H) \simeq 2.5 \times 10^{-7} \textnormal {GeV}
\end{eqnarray}
For the $1.7 \times 10^7 $ Z decays accumulated by the ALEPH, DELPHI, L3 and OPAL experiments at the LEP \cite{ALEPH:2005ab}, this gives an expectation of only about two such events. As in our model, the mass of the dark matter candidate $H$ is not zero but can be higher (but lighter than $m_h$), thus event expectation can be higher and it may have some prospects of detection at the proposed $e^+e^-$ detector ILC. \\
The phenomenology of the pseudoscalar $A$ and the charged scalars $H^{\pm}$ are similar in nature. 
The charged scalar particle $H^{\pm}$ can be produced via Drell-Yan process and it will be quite elusive to observe at the LHC. As the mass of the heavier charged scalar $m_H^\pm$ is greater than the mass of the neutral scalar $m_H$($H$ being lightest in the scalar mass spectrum), $H^{\pm}$ can decay to $H^{\pm} \rightarrow W^{\pm} H$. It will be quite elusive to discover $H^\pm $ with the decay products of  "2 W-boson + Miss $E_T$" final states at the LHC. If the mass of the pseudoscalar $A$ is less than the mass of the charged scalar $H^\pm$, $H^{\pm}$ can also decay to $H^{\pm} \rightarrow W^{\pm} A$. On the other hand, the pseudoscalar $A$ (depending on the actual mass hierarchy) can also decay to the charged scalar $H^{\pm}$ in association with a $W^{\pm}$ if $m_A > m_H^\pm$. In the case of $m_{H^{\pm}} > m_w + m_{H/A}$, the expressions for the partial decay widths of the heavy charged scalar $H^{\pm}$ are given by,
\begin{eqnarray}
\Gamma({H^{\pm}} \to W^{\pm} H)&=&\frac{g^2}{64 \pi M_w^2 M_{H^{\pm}}^3}\left(( M_{H^{\pm}}^2 - M_{H}^2 - M_w^2)^2 - 4 M_{H}^2 M_w^2 \right)^{3/2},\\
\Gamma({H^{\pm}} \to W^{\pm} A)&=&\frac{g^2}{64 \pi M_w^2 M_{H^{\pm}}^3}\left(( M_{H^{\pm}}^2 - M_{A}^2 - M_w^2)^2 - 4 M_{A}^2 M_w^2 \right)^{3/2}
\end{eqnarray}  
In this section we have given a reasonable idea of the possible productions and decays of the five scalar particles $h, H ,A$ and $H^\pm$. The neutral scalar $h$ has the same couplings as the newly discovered $125$ GeV SM-like Higgs boson and will have the exact same phenomenology. The other neutral scalar $H$ does not couple to the fermions at all and is also the lightest scalar in the mass spectrum. $H$ can be a DM candidate as it is the lightest $Z_2$ negative particle. The bounds on its mass $m_H$ and the couplings with the SM-like Higgs $h$ coming from the invisible Higgs branching ratio upper bounds and Electroweak precision parameters have been discussed here. The charged scalars $H^\pm$ and the pseudoscalar $A$ also do not couple the fermions and the related phenomenology has also been discussed briefly.
\section*{Summary and Conclusions}
We have presented a simple twist in the well studied two Higgs doublet model in the form of adding an extra interchange symmetry between the two Higgs doublets ($\Phi_1$ $\leftrightarrow$ $\Phi_2$). A residual $Z_2$ symmetry that remains unbroken after the original symmetry $\Phi_1$ $\leftrightarrow$ $\Phi_2$ is spontaneously broken makes the charged scalars $H^\pm$, the neutral scalar $H$ and the pseudoscalar $A$ to have  $Z_2$ negative charges and all the other fields remain $Z_2$ positive. This in turn makes the lightest $Z_2$ negative neutral scalar $H$ to be the Dark Matter candidate. This neutral scalar $H$ can be much lighter in mass than the Standard Model-(SM) like neutral scalar $h$ with mass $m_h \simeq 125$ GeV as seen by the LHC. Interestingly this lighter neutral scalar $H$ as well as the charged scalars $H^\pm$ and the pseudoscalar$A$ do not couple to fermions.The lighter neutral scalars also don't have the usual three point couplings with the Gauge bosons($W^\pm$ and $Z$) present in the Standard Model; but only has four-point couplings with $W^\pm$ and $Z$. The only way to produce the lightest $Z_2$ negative DM candidate $H$ is through the decays of the SM-like neutral scalar $h$ where this SM-like neutral scalar $h$ will have an extra invisible decay channel through $h \rightarrow H H$. The Invisible decay branching ratio of the $125$ GeV SM-like Higgs can be as large as ${Br_{inv}}_h < 25 \%$. We study the parameter space of the effective coupling $\lambda^*$ between the neutral scalars ($hHH$) and the mass of the DM candidate lighter neutral scalar $m_H$. We also comment on the electroweak constraints in this scenario and also the other possible phenomenology for the charged scalars $H^\pm$ and pseudoscalar $A$.

\begin{acknowledgments}

SC is thankful to S. Nandi for useful discussions and communications. SC would like to thank Colby College and Fermilab for their hospitality where part of this work was completed. H. B would like to thank Colby College and Yale University for providing the computational resources used in this work. This research was supported in part by the Colby College Natural Sciences Research Grant 2018-2019.

\end{acknowledgments}


\end{document}